# Symmetry Partition Sort[*]
## (Preliminary Draft)

Jing-Chao Chen
Email: chen-jc@dhu.edu.cn
*School of informatics, DongHua University*
*1882 Yan-an West Road, Shanghai, 200051, P.R. China*
(June 1, 2007)

**SUMMARY**

In this paper, we propose a useful replacement for quicksort-style utility functions. The replacement is called Symmetry Partition Sort, which has essentially the same principle as Proportion Extend Sort. The maximal difference between them is that the new algorithm always places already partially sorted inputs (used as a basis for the proportional extension) on both ends when entering the partition routine. This is advantageous to speeding up the partition routine. The library function based on the new algorithm is more attractive than Psort which is a library function introduced in 2004. Its implementation mechanism is simple. The source code is clearer. The speed is faster, with $O(n \log n)$ performance guarantee. Both the robustness and adaptivity are better. As a library function, it is competitive.

## INTRODUCTION

It has been all along one of the studied problems how to devise a better library sort function. A large amount of library sort functions have been proposed. Most of them were derived from Quicksort [1] due to Hoare. The Seventh Edition [9] and the 1983 Berkeley function in Unix are the early library function based on Quicksort. The late one is Bentley and McIlroy's Quicksort (BM qsort, for short) [2] proposed in 1993. Whether the early one or the late one, in theory, they consume possibly quadratic time [2,11].

Let $W(n)$ and $A(n)$ denote the number of comparisons in the worst case and on the average, respectively. Table 1 shows the number of comparisons required by some practical sorting algorithms. The Psort [5] shown in Table 1 is a library function recently developed by Chen, which is based on Proportion Extend Sort (PEsort, for short) [3]. Although Psort is a variant of PEsort, no theoretical evidence shows that the two algorithms have the same time behavior. This is still an open problem. Therefore, it is nothing but conjecture that $W(n)$ of Psort is $\theta(n \log n)$. Full Sample Sort (Fsort, for short) [7] is also a practical algorithm, which is of some features similar to PEsort. Like PEsort, its number of comparisons both in the worst case and on the average is bounded by $O(n \log n)$ [3,7,8]. Their worst case numbers of exchanges both are bounded by $O(n \lg^2 n)$ [7 8]. In practice, both Psort and Fsort are faster, more robust and more adaptive than functions based on Quicksort. However, in some particular cases, they are not the fastest. For example, for sorting random integers, the best-of-three version of Quicksort with an extra storage [10] could run faster than they. Also, their code is not very clear. This needs to be improved further.

Table 1. The number of comparisons

| Algorithms | $W(n)$ | $A(n)$ | References |
|---|---|---|---|
| Proportion Extend Sort | $\theta(n \log n)$ | $n \log n + O(n)$ | 3, 7, 8 |
| Psort | unknown | $n \log n + O(n)$ (observed) | 5 |
| Full Sample Sort | $\theta(n \log^2 n)$ | $n \log n + O(n)$ | 7 |
| BM qsort | $\theta(n^2)$ | $1.1n \log n + O(n)$ (observed) | 2, 11 |

The paper introduces a new algorithm that has the same time complexity as PEsort. But the algorithmic mechanism is different. The new algorithm is referred to as Symmetry Partition Sort. Its symmetry is mainly reflected in the partition portion. When entering the partition routine, it always ensures that the already sorted small and large items are collected in two ends of the items to be split, respectively, so that the partition routine needs not consider the border test of index variables. The advantage of this partition scheme is that it can simplify the "fat partition" routine, and outperform the best-of-three version of

[*] This work was partially supported by the National Natural Science Foundation of China grant 60473013

Quicksort in any case, even in a case favourable to Quicksort. Compared with the library function Psort, the code of this algorithm is clearer. To speed up the algorithm, besides using some traditional tricks such as sorting small array by insertion sort, we remove the tail recursions. To enhance adaptive performance, we get the longest initial sorted sequences by extracting the sorted items prior to entering the sort. Empirical results show that the resulting library function is faster and more robust than both Psort and Fsort. In terms of *Rem*-adaptivity (its definition will appear below), it has good behavior. In some cases, it can achieve even the adaptive performance of McIlroy's Mergesort with Exponential Search (MSES for short) [13]. Like PEsort, it can be guaranteed theoretically to require $O(n \log n)$ comparisons in the worst case, and $n \log n + O(n)$ comparisons on the average case. Symmetry Partition Sort is also an algorithm with Quicksort flavour. Consequently, it has also excellent caching behaviour like Quicksort.

## THE ALGORITHM

Symmetry Partition Sort is based on PEsort. First we review briefly PEsort before describing Symmetry Partition Sort. In PEsort, a sort is considered as the sort of such a partially sorted array: the portion to the left, $S$, is sorted; the remainder, $U$, is unsorted. Below we describe a call $(S, U)$ to PEsort in a pseudo-code.

Define $V$ as
$$V = \begin{cases} \text{the subarray of size } p|S| \text{ to the left of } U & \text{if } \beta|S| \leq |SU| \\ U & \text{otherwise} \end{cases}$$
where $p$ and $\beta$ are a constant.

We sort a partially sorted subarray $(S, V)$ in such a way:

Partition $V$ into $V_L$ and $V_R$ by the median $m$ of $S$ such that $\max(V_L) < m < \min(V_R)$.
Let $S_L$ be the left half of $S$, $S_R$ the right half of $S$. Namely, $S$ is viewed as $S_L m S_R$.
Swap the blocks $mS_R$ and $V_L$, preserving the order of items in $mS_R$
Sort recursively the subarrays $(S_L, V_L)$ and $(S_R, V_R)$, respectively.
Finally, we sort array $(S', U - V)$ recursively, where $S'$ is the sort of $SV$.

Unlike PEsort, Symmetry Partition Sort can complete the sort of not only left partially sorted arrays, but also right partially sorted arrays. Here, an array is said to be a left partially sorted array if the portion to its left is sorted. Similarly for a right partially sorted array. The algorithm assumes that the input is of the form

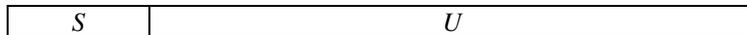

or

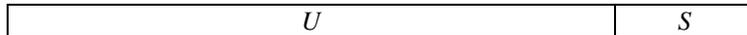

where $S$ and $U$ denote the sorted and unsorted subarray, respectively. By a vector left or right move, we transform it into four:

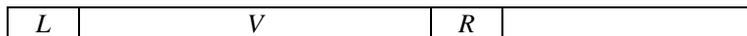

where $LR = S$ and the size of $V$ is defined as before. Using the last item $m$ of $L$, i.e., the median of $S$, we partition $V$ into two parts to yield:

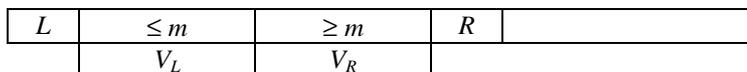

Now sort recursively the subarrays $(L, V_L)$ and $(V_R, R)$ to get the state with a longer sorted sequence $S'$ like

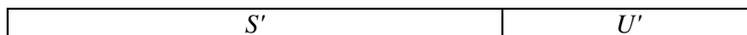

Finally, we sort recursively the array $(S', U')$ to complete the sort of the whole array. Next we formalize the algorithm in a pseudo-code.

**procedure** SymmetryPartitionSort $(S, U, sign)$
    **if** $U$ is empty **then return**
    Let $S = LR$ with $|L|=|S|/2$
    **if** $sign = -1$ **then** transform $US$ to $LVRU'$ by vector left move operations
    **else** transform $SU$ to $LVRU'$ by a vector right move operation
    Partition $V$ into $V_L$ and $V_R$ by the last item $m$ of $L$ such that $\max(V_L) \leq m \leq \min(V_R)$
    SymmetryPartitionSort $(L, V_L, 1)$
    SymmetryPartitionSort $(R, V_R, -1)$
    SymmetryPartitionSort $(S', U', 1)$, where $S'$ is the output of the previous two recursive calls
**end procedure**

The parameter *sign* is used to indicate whether the sorted subarray $S$ is to the left of $U$ or to the right of $U$. That is, $S$ is to the right of $U$ when $sign = -1$, and to the left of $U$ otherwise. In the concrete implementation, parameters $S$ and *sign* can be merged into one parameter. Let $s$ be the size of the sorted subarray plus a sign. When $S$ is to the left of $U$, $s$ is set to $|S|$. And when $S$ is to the right of $U$, $s$ is set to $-|S|$. The significant difference between PEsort and Symmetry Partition Sort is that the former completes the block swap after partitioning $V$, while the latter does prior to partitioning $V$. This is advantageous to speeding up the partition routine. When partitioning $V$, both ends of $V$ are occupied by the sorted items. Thus, we can view the last sorted item to the left and the first sorted item to the right as sentinels. The end-sentinel partitioning is more efficient than the other partitioning, since it can ignore the computation on whether an index reaches the end of the array. Let $L = a[0..v1]$, $V = a[v1+1..v2]$, $R = a[v2+1..r]$ and the function of swap(a+i, b+j) be to exchange the values in a[i] and b[j]. The following is the C code for partitioning $V$ with a sentinel-end partitioning technique.

```
i =v1+1; j = v2;
do {    while ( a[i] < a[v1] ) i++;
        if (i >= j ) break;
        while ( a[j] > a[v1] ) j--;
        if( i >= j ) break;
        swap(a+i, a+j); i++; j--;
} while ( i <= j );
```

If $L = \phi$ or $R = \phi$, this partitioning technique will fail. However, this can be avoided easily, since we can create the sorted subarray $S$ with three items by sorting the first, middle, and last items, and setting the first, middle items to $L$ and last item to $R$. Using this technique, this algorithm can outperform the best-of-three version of Quicksort with an extra storage even on random integers. The symmetry of this algorithm is mainly reflected in this portion.

## VARIOUS OPTIMIZATIONS

The algorithm given in the previous section is the basic version of Symmetry Partition Sort. Directly implementing it will detract its many practical merits. Therefore, we improve on it as follows.

1. Sort a small array by insertion sort. This trick has been used widely by BM qsort, Psort etc. For its implementation details, see Appendix and References [2] and [5].
2. Speed up the algorithm on nonrandom inputs by a uniform sampling technique.
3. To avoid slowdown efficiently on some reasonable inputs, such as many identical items, we use the same partitioning scheme as Psort. Its goal is to divide a sequence three subsequences: small items, equal items and large items. For details, see Appendix and Ref. [5].
4. Eliminate the tail recursions.
5. Tune the values of $p$ and $\beta$. Based on empirical results, the optimal value of $p$ is 16. The optimal value of $\beta$ is 512 when $|S| \leq 2$, and 256 otherwise.
6. Create the longest possible sorted subarray as the initial inputs. This strategy is used to improve the adaptivity of the algorithm.

Below we expand the details of some tricks above-mentioned.
   The core of our uniform sampling technique is to extract sampling items evenly (at equal space intervals) from the unsorted subarray $U$, and place them into $V$ prior to sorting the partially sorted subarray $LV$. The sampling space is set to n/v, where n is the number of items to be sorted, and v =$|V|$. Let $U = a[u1..u2]$. The following is a C program for extracting v − u1 sampling items.

```
int  i, j;
for(i = j = u1; i < v; i++, j = j + n/v) swap(a+i, a+j);
```

The timing to invoke the program depends on whether the input satisfies the balanced (random) property. This is done by testing if $|U| = |V|$ or not. That is, whenever $|U| \neq |V|$ (this is considered as a nonrandom input), we call this program. Incorporating this technique can beat the number of comparisons down to $n \log n$ for sorted inputs.
   As has been seen before, the algorithm is described in a recursive way. In general, a recursive call needs maintain some local variables. Nevertheless, this is not cheap. However, implementing the Symmetry

Partition Sort in a full non-recursive way is also expensive. Based on our observation, we found that a hybrid fashion is more efficient than any pure fashion. In the realistic implementation, we replace the last recursive call with an unrolling way, keep other recursive calls unchanged. In the appendix, one will see that the third recursive call SymPartitionSort(a, v, n, es, cmp) is replaced with the structure:

```
while (1){
       ⋮        ⋮        ⋮
    s=v;
}
```

When v = n, the second recursive call becomes the last recursive call. At this time, this recursive call is removed also. For details, see Appendix. Additionally, in order to reduce the number of recursive calls, we add a test to ask whether the set of the unsorted items is empty before each recursive call.

A non-random sequence is often one formed by interleaving a sorted subsequence and a random subsequence. This is a common phenomenon in updating data bases. How to speed up the sort on this class of inputs is the problem we address here. The feature of the algorithm is that the longer the initial sorted subarray, the fewer the number of comparisons required. This is similar to the algorithm in [6]. Therefore, we decide to use a similar way to [6] to collect sorted items together to the left end as an initial input, and then invoke Symmetry Partition Sort. The resulting algorithm will be adaptive. The collection proceeds as follows.

(1) Find the first run (we say that a subsequence is a run if it is ordered (increasing or decreasing)).
(2) Find the next run.
(3) Concatenate the current run and its left run it into a longer run.
(4) Repeat (2) and (3) until all items are scanned.

This is actually a simplified "natural mergesort". Problems to be addressed here are how to find a required run and how to concatenate two runs efficiently. As a possible solution to the first problem, we take the $1^{st}$ item as the start of a run, and then find its tail from left to right. If the length of the run found is not sufficiently long, say less than 4, we abort it, and then re-find a run with the $2^{nd}$ item as a start. The process is repeated until a run is found, or all items are scanned. This method is simple but inefficient. For this reason, we modify it as follows. Take some item (which is 10 items away from the previous run) as the middle of the series to be scanned, find backward the start of a run, and find forward the tail. If the run length is not enough, we abort it, take the next 10-th item as the start of a new scan, and resume the process.

We concatenate two runs together by a head-tail-elimination strategy. The first step to find two knots where two runs can be tied in a longer run by scanning backward the $1^{st}$ run and forward the $2^{nd}$ run. The second step is to bring the right portion (from the $2^{nd}$ knot to the end) of the $2^{nd}$ run to the right end of the $1^{st}$ knot to obtain a longer run. For example, given the $1^{st}$ run=<1,3,5,20,21> and the $2^{nd}$ run=< 2,4,9,10,11>, then the $1^{st}$ and $2^{nd}$ knots found are items 5 and 9. The merged longer run is <1,3,5, 9,10,11>.

The preprocess for extracting a sorted sequence from an input is harmful to a random input. To reduce this harm, we detect the nature of an input by counting inversions and non-inversions on a sample set, which is formed by taking an item every 97 (this is an empirical result) items. Given an input a[0..n-1]. The following C code is to compute the difference D_inv of inversions and non-inversions in array a.

```
for (D_inv=0, i=0; i < n; i+=97)  if ( a[i-1] < a[i] ) D_inv++; else D_inv--;
```

If |D_inv| is small, say |D_inv| < n/512, the input is possibly random. In this case, we abort the preprocess. Otherwise, we continue to the search and merge of runs.

Based on the above ideas, the resulting adaptive algorithm may be summarized in a pseudo-code as follows.

**procedure** AdaptiveSymmetryPartitionSort (*A*[1.. *n*])
    find the $1^{st}$ run *S*
  **if** the input *A* is a non-random **then**
      *i* ← (pointer to the tail of *S*) +10
    **while** *i* < *n*  **do**
        find next run *T* backward and forward from the *i*-the item in *A*
        merge *S* and *T* into *S* by head-tail-elimination strategy
        *i* ← (pointer to the tail of *T*) +10
    **end while**
  **end if**
  **if** *S* is reverse **then** reverse(*S*)
  SymmetryPartitionSort (*S*, *A* − *S*, 1)
**end procedure**

The appendix presents the C codes of an optimized Symmetry Partition Sort and an adaptive Symmetry Partition Sort, which support an interface identical to the C library function qsort. They are named as SymPartitionSort and Adp_SymPsort, respectively.

## EMPIRICAL STUDIES

In this paper, we conducted the experiments about running times with two personal computers: Celeron CPU 700MHz and Pentium IV CPU 2.66GHz. All the programs were coded in C and adopted an interface identical to the C library sort function sort (char * A, int s, int n, int length, int (cmp)( const void *, const void * )), where A points to the start of an array to be sorted; n is the number of items in A ; length is the size in bytes of each item; cmp is a comparison function which returns an integer that is negative, zero or positive, depending on the comparing result of its two arguments.

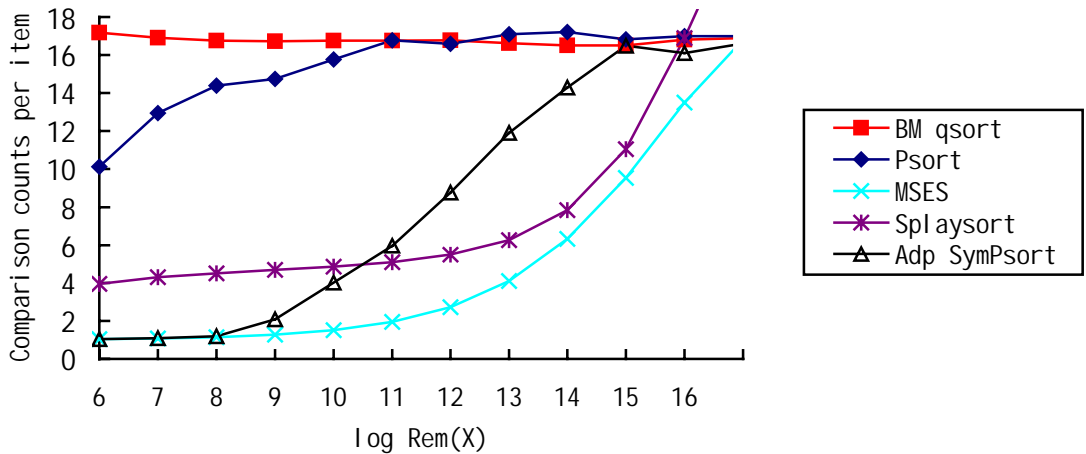

(a) Nearly sorted inputs

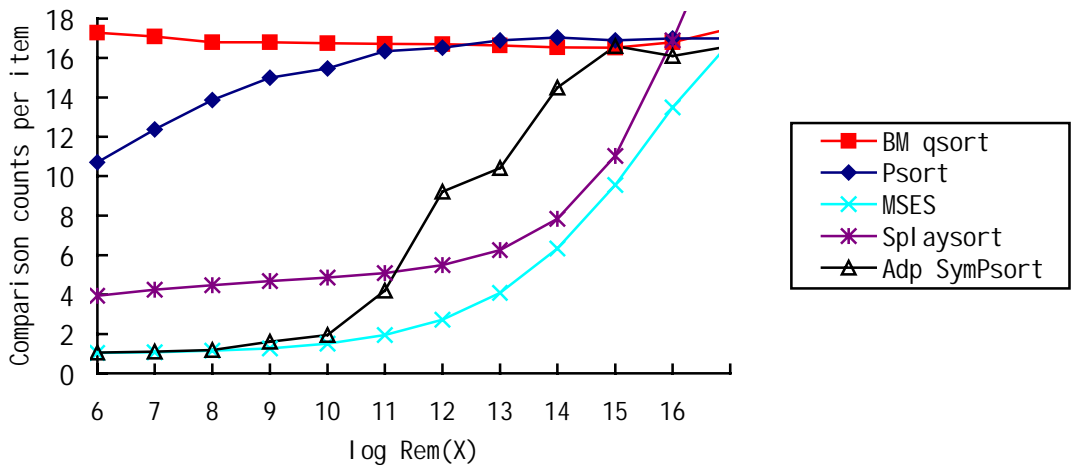

(b) Reverse nearly sorted inputs

Figure 1. *Comparison counts per item to sort a non-random input X of n =100000 items, plotted as a function of* log *Rem(X)*

We evaluated the proposed sorting algorithm with such performance metrics as *adaptivity, running time, comparison count, exchange count*, *robustness* etc. Various measures of presortedness have been proposed. An important measure is *Rem*, which is defined as the number of items that must be eliminated to leave a sorted sequence. Let *X* is some *n*-sequence to be sorted. The definition is formalized as

$Rem(X) = n - \max\{k : X \text{ has an ordered subsequence of size } k\}$.

The lower the value, the more ordered the sequence. $Rem(X) = 0$ implies that the sequence $X$ is a sorted or reverse sorted one. In real data applications, it is one of the most appealing measures.

We generate a nearly ascending sequence $X$ with $Rem(X) = k$ in such a way: create first the ascending $x_1 < x_2 < ,…,< x_n$, and then randomly choose $k$ indexes and replace integers pointed to by them with $k$ random integers such that for each replaced integer $x_{i+1}$, $x_i > x_{i+1}$ or $x_{i+1} > x_{i+2}$. In a similar way, we generate a nearly descending sequence $X$ with $Rem(X) = k$.

Figure 1 shows average comparison counts per item required to sort an incompletely shuffled input of $n=100000$ items as a function of log $Rem(X)$ for five sorting algorithms. Throughout this paper, we provided 20 different inputs for each problem. Each algorithm was run on the same collection of inputs. Adp SymPsort refers to the adaptive Symmetry Partition Sort given in the Appendix. Contrasting Figure 1 (a) to Figure 1 (b), we noted that these algorithms, on nearly sorted inputs or on reverse nearly sorted inputs, have the same behaviour with respect to *Rem*. For small values of log $Rem(X)$, Adp SymPsort achieves the same *Rem*-adaptivity as Merge Sort with Exponential Search (MSES for short)[†] due to McIlroy [13]. For large values of log $Rem(X)$, the situation is a little different. The best was still MSES, but the second best was Splaysort[††] [12], followed by Adp SymPsort. BM qsort is not *Rem*-adaptive in any way.

Figure 2 shows average times (ms) taken to sort an input $X$ of 100000 items on Pentium IV 2.66GHz as a function of log $Rem(X)$. Each item contained a 4-byte pointer to an 8-byte floating integer used as a key. Unexpectedly, Adp SymPsort was almost consistently faster than MSES in this experiment. When the order of items is close to randomization, MSES was slower than even BM qsort. Splaysort run very slow also, especially for large values of $Rem(X)$. This is why both MSES and Splaysort have poorer cache performance and heavy data movements so that they were very slow in many cases, especially for random inputs.

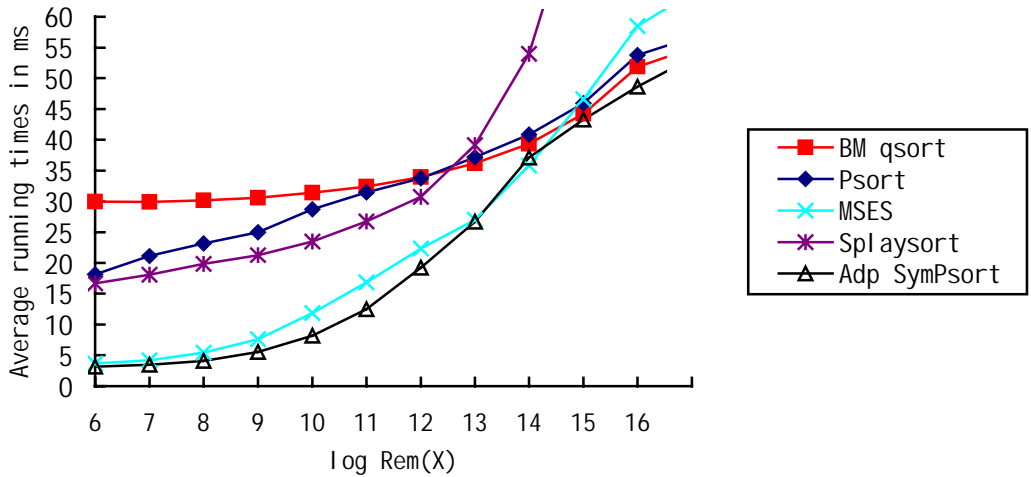

Figure 2. *Time in microseconds to sort an incompletely shuffled input X of n =100000 4-byte pointers to 8-byte floats, plotted as a function of* log *Rem(X)*

An another appealing measure of presortedness is *Inv*, which is defined as the number of pairwise inversions,

$$Inv(X) = |\{(i, j) : 1 \le i < j \le n \text{ and } x_i > x_{i+1}\}|$$

To conduct experiments with *Inv*, we generated sequences with $kn/2$ or fewer inversions in such a way: Starting with a sorted sequence, the $n$ items were divided into blocks of size $k$ and the items in each block were permuted so that each block had at most $k^2/2$ inversions and the total $Inv(X)$ was less than $kn/2$. With respect to *Inv*, Adp SymPsort has the same performance as BM qsort. Both are not *Inv*-adaptive. Figure 3 shows average comparison counts per item required to sort an incompletely random input $X$ of $n=100000$ items as a function of log $(2Inv(X)/n)$. This measure had a significant impact on Psort, but not on Adp SymPsort. Except for $Inv(X)<n/2$, Adp SymPsort made the same comparisons as on the completely random input, and was a little better than BM qsort. This shows that Adp SymPsort is still robust.

---

[†] The C code for MSES is available from http://gcov.php.net/PHP_5_2/lcov_html/main/mergesort.c.gcov.php
[††] The C code for Splaysort is available from http://www.cs.mu.oz.au/~alistair/splaysort.c

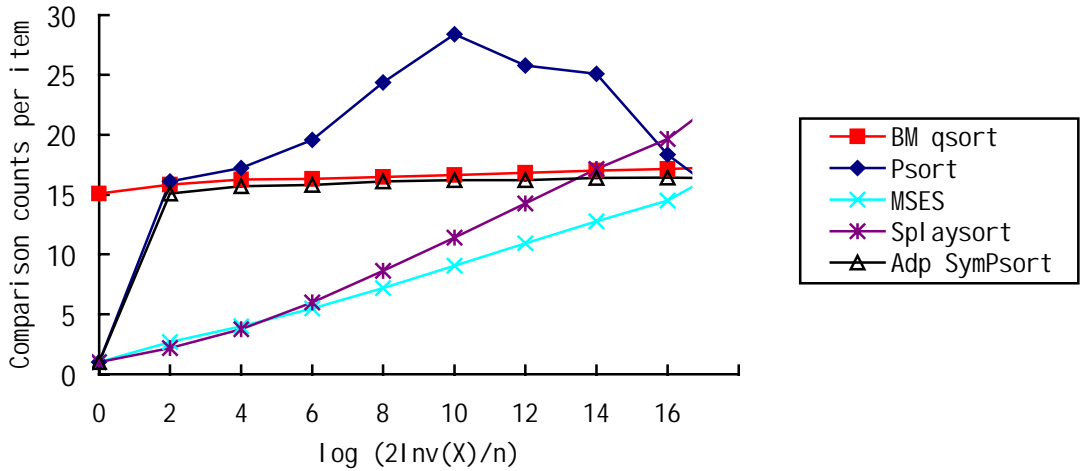

Figure 3. *Comparison counts per item to sort an input sequence X of n =100000 items, plotted as a function of* log k, *where* 2Inv(X) < kn.

Table 2 shows the average time (ms) taken to sort *n* 4-byte integers on Pentium IV 2.66GHz. In this experiment, the key comparing was set to be the following typical integer comparison function

    int cmp(const void * i, const void *j) { return *(int *)i - *(int *)j; }

which is the cheapest. In Table 2, column, distinct, presents the time taken to sort *n* randomly permuted distinct integer, while column, mod *n*, presents the time taken to sort *n* random integers taken modulo *n*. Algorithms Adp Psort and Adp Fsort are one obtained by replacing SymPartitionSort in Adp_SymPsort with Psort [5] and Fsort [7], respectively. Besides this substitution, we eliminate the adaptive process within Psort to speed up Adp Psort. In Figures 1~3, we did not list the empirical results of Adp Psort and Adp Fsort, since they had the same adaptivity as Adp SymPsort. Whether on distinct integers or integers taken modulo *n*, Adp SymPsort was the fast, and 3 ~ 5 per cent faster than BM qsort. An interesting phenomenon is that Adp SymPsort was always faster in case "mod *n*" than in case "distinct", while Adp Psort and Adp Fsort was converse. This may be due to that shifting the sorted subarray prior to partitioning favors simplifying the subsequent equal-items movement. Because MSES is not a competitor for random inputs, Table 2 does not give its empirical result, for which, the interested reader is referred to [5]. The reason why Psort was not chosen as a reference point is that the performance of Adp Psort is better than that of it.

Table 2. The average time in ms to sort *n* integers (Pentium IV 2.66GHz)

| n | Adp SymPsort | | Adp Psort | | BM qsort | | Adp Fsort | |
|---|---|---|---|---|---|---|---|---|
| | distinct | mod *n* | distinct | mod *n* | distinct | mod *n* | distinct | mod *n* |
| 25000 | 5.884 | 5.777 | 6.103 | 6.131 | 6.155 | 6.068 | 6.050 | 6.104 |
| 50000 | 12.49 | 12.28 | 12.96 | 13.01 | 13.11 | 12.90 | 12.87 | 12.97 |
| 100000 | 26.61 | 26.16 | 27.51 | 27.59 | 27.81 | 27.40 | 27.32 | 27.49 |
| 500000 | 153.25 | 150.90 | 155.93 | 156.73 | 157.98 | 155.95 | 155.93 | 158.18 |
| 1000000 | 320.05 | 316.45 | 327.40 | 329.10 | 331.05 | 327.45 | 331.60 | 333.80 |
| 2000000 | 672.30 | 665.30 | 686.40 | 689.30 | 693.60 | 687.00 | 695.80 | 700.20 |

As shown in Table 3, in terms of the number of comparisons, the fewest was Adp SymPsort, which was consistently better than other three algorithms Adp Fsort, Adp Fsort and BM qsort. In terms of the number of swaps, it was a little poorer than the three algorithms. This is reflected in Table 4. The model where swapping is expensive does not favors Adp SymPsort. This can be avoided by reducing the number of swaps with a "ripple swap" trick introduced in [4]. Though, in the general case, this strategy is not necessary. Table 5 shows the average time (ms) taken to sort 100000 distinct items with each key a 4-byte integer for different item sizes on Celeron 700MHz and Pentium IV 2.66GHz. The purpose of this experiment is to gauge the influence of movement operations on the performance. When the number of bytes in each item is less than or equal to 9, Adp SymPsort, whether on Celeron 700MHz or Pentium IV 2.66GHz, was the fast. As the number of bytes in each item increases, its advantage disappears. When the number of bytes per item is equal to 13, Adp SymPsort was slower than Adp Psort on Pentium IV 2.66GHz.

However, if each item is word-aligned, swapping becomes cheap, so that Adp SymPsort was the fast until the number of bytes per item reaches 52. Celeron 700MHz seems to advantage Adp SymPsort, since the result reported in Table 5 is that it was always faster than other algorithms on this machine, but was not so on another machine.

Table 3. The average number of comparisons on $n$ distinct random items

| $n$ | Adp SymPsort | Adp Psort | BM qsort | Adp Fsort |
|---:|---:|---:|---:|---:|
| 1000 | 9653 | 10305 | 9952 | 9941 |
| 25000 | 355220 | 367980 | 375074 | 360364 |
| 50000 | 766785 | 786760 | 803076 | 770541 |
| 100000 | 1639944 | 1688635 | 1718466 | 1641247 |
| 500000 | 9267453 | 9538272 | 9861301 | 9401153 |
| 1000000 | 19703924 | 20168360 | 20871158 | 19799318 |
| 2000000 | 41441219 | 42532145 | 43934867 | 41606187 |

Table 4. The average number of swaps on $n$ distinct random items

| $n$ | Adp SymPsort | Adp Psort | BM qsort | Adp Fsort |
|---:|---:|---:|---:|---:|
| 1000 | 3201 | 2781 | 2835 | 2832 |
| 25000 | 110155 | 100413 | 98891 | 102405 |
| 50000 | 227656 | 211967 | 210072 | 218161 |
| 100000 | 479511 | 445392 | 443949 | 463513 |
| 500000 | 2768386 | 2562361 | 2502998 | 2629568 |
| 1000000 | 5651452 | 5321372 | 5243871 | 5530168 |
| 2000000 | 11810531 | 11078900 | 10975299 | 11604469 |

Table 5. The average time in ms to sort 100000 distinct items

| Bytes per item | Adp SymPsort | | Adp Psort | | BM qsort | | Adp Fsort | |
|---:|---:|---:|---:|---:|---:|---:|---:|---:|
| | 700MHz | 2.66GHz | 700MHz | 2.66GHz | 700MHz | 2.66GHz | 700MHz | 2.66GHz |
| 9 | 148.8 | 45.07 | 149.9 | 45.12 | 153.1 | 45.56 | 152.7 | 46.41 |
| 13 | 191.8 | 54.14 | 194.6 | 53.10 | 201.1 | 54.15 | 196.4 | 58.58 |
| 14 | 202.6 | 55.94 | 203.5 | 54.63 | 207.8 | 55.86 | 207.2 | 60.32 |
| 52 | 535.3 | 57.20 | 538.5 | 57.38 | 544.5 | 59.16 | 547.8 | 58.82 |
| 80 | 763.0 | 83.38 | 764.5 | 82.48 | 763.0 | 85.20 | 788.0 | 83.72 |

Table 6. The average number of comparisons on $n$=100000 random integers taken modulo $k$

| $k$ | Adp SymPsort | Adp Psort | BM qsort | Adp Fsort |
|---:|---:|---:|---:|---:|
| 2 | 150995 | 151528 | 150077 | 151062 |
| 10 | 291261 | 296038 | 306123 | 291146 |
| 100 | 588794 | 589991 | 622491 | 583492 |
| 1000 | 941066 | 940736 | 982405 | 927783 |
| 10000 | 1309552 | 1330522 | 1368386 | 1297330 |
| 100000 | 1577997 | 1616587 | 1645520 | 1566716 |

Each algorithm listed in Tables 6 and 7 contained the fat partition technique, which is used to improve the repeated key adaptivity of algorithms. As shown Tables 6 and 7, Adp SymPsort was the best with respect to repeated key adaptivity, and took not only fewer comparisons, but also fewer swaps, on the average in almost all cases, except for modulo $n$, in which, it took more swaps. In fact, the case modulo $n$ is almost identical to that of distinct random items, since in the case there is almost no equal item.

Ref. [5] showed a testbed for certifying performance, which was introduced first by Bentley and McIlroy [2]. The testbed is very representative and covers broadly various plausible inputs such as sawtooth, random, flat, shuffled, dither, sorted, reverse sorted etc. For its detailed pseudocode, please see Figure 1 in Ref. [5]. Here we used also it to certify the robustness of algorithms. Table 8 shows the worst case number of comparisons required to sort $n$ items on the testbed. For different $n$, the number of input instances generated is different, and varies from 510 to 660. The algorithm, Adp SymPsort, proved empirically to be the most robust. The number of comparisons never exceeded $1.2n \log n$. The worst case was consistently better than that of BM qsort. The cases in which the number of comparisons exceeds $1.1n \lg n$ were fewer than 0.6%. Table 9 shows the worst case number of swaps required to sort $n$ items on the testbed. In the worst case, the number of swap required by our algorithm was more than BM qsort, and fewer than Adp Psort. But the number never exceeded $0.45n \log n$.

Table 7. The average number of swaps on $n$=100000 random integers taken modulo $k$

| $k$ | Adp SymPsort | Adp Psort | BM qsort | Adp Fsort |
|---|---|---|---|---|
| 2 | 64674 | 114313 | 149857 | 141865 |
| 10 | 148946 | 196379 | 194200 | 207636 |
| 100 | 230114 | 272776 | 268262 | 293337 |
| 1000 | 312837 | 335710 | 347476 | 379142 |
| 10000 | 394765 | 397514 | 421214 | 455642 |
| 100000 | 462840 | 436790 | 434450 | 461609 |

Table 8. The worst case comparisons on the testbed shown in Fig. 1 of Ref. [5]

| $n$ | Instance # | The worst case number of comparisons | | | | % over $1.1n \log n$ | | | |
|---|---|---|---|---|---|---|---|---|---|
| | | Adp SymPsort | Adp Psort | BM qsort | Adp Fsort | Adp SymPsort | Adp Psort | BM qsort | Adp Fsort |
| 50000 | 510 | 916020 | 913430 | 920167 | 912099 | 0.6 | 2.2 | 2.9 | 0.2 |
| 1000000 | 630 | 22933715 | 26369317 | 23927787 | 23504806 | 0.5 | 1.9 | 3.5 | 0.8 |
| 2000000 | 660 | 45327892 | 55614163 | 49855553 | 49549883 | 0 | 1.7 | 3.8 | 0.8 |

Table 9. The worst case number of swaps on the testbed shown in Fig. 1 of Ref. [5]

| $n$ | Instance # | Adp SymPsort | Adp Psort | BM qsort | Adp Fsort |
|---|---|---|---|---|---|
| 50000 | 510 | 343656 | 351922 | 217260 | 322076 |
| 1000000 | 630 | 8877398 | 8950155 | 5302093 | 8439380 |
| 2000000 | 660 | 18695152 | 18895316 | 11061038 | 17823754 |

## CONCLUSIONS

In the paper, we have presented a practical sorting algorithm as an alternative to PEsort. Surprisingly, this algorithm has not only simpler implementation, but also better performance. A series of experiments proved that this algorithm is robust, efficient and equally cache-friendly to Quicksort for an in-cache phase of cache-aware sorting. How to use efficiently cache in modern machines has received extensively attention [14]. In the sense, this algorithm should be useful. Although theoretically this algorithm has a case of $n \log^2 n$ exchanges as PEsort [8], this can be easily avoided, if we use an in-place merging strategy when the size of the sorted subarray is greater than that of the unsorted subarray. Is it necessary to add the strategy to this algorithm? This will be a problem deserving of studying.

## ACKNOWLEDGEMENTS

The author is grateful to A. Mycroft and M. Richards for their helpful discussions while visiting University of Cambridge.

## APPENDIX: THE C CODE OF ADAPTIVE SYMMETRY PARTITION SORT

The C source code listed below is available from
http://cist.dhu.edu.cn/professors/AdpSymmetryPSort.cpp
Before the C code for Adaptive Symmetry Partition Sort, we give several macros. The swap strategies of items depend on the value of variable swaptype, which is set by the following macro:

```
#define SWAPINIT(a,es) swaptype =                    \
 (a - (char *) 0) % sizeof(long) || es % sizeof(long) ? 2 : \
 es == sizeof(long) ? 0 : 1
```

Using the variable swaptype, we classify three kinds of swaps: zero for swapping single long integers, one for multiple long integers, and two for strings of chars.
  Swapping two vectors, each of which is a sequence of n bytes which are aligned, is done by the following macro:

```
#define swapvector(TYPE,pi,pj,n)          \
```

```
        do {                                           \
            TYPE t= *(TYPE *) (pi);                    \
            *(TYPE *) (pi) = *(TYPE *) (pj);           \
            *(TYPE *) (pj) = t;                        \
            pi+=sizeof(TYPE); pj+=sizeof(TYPE);        \
            n-=sizeof(TYPE);                           \
        } while (n > 0);
```

An efficient function for swapping two n-byte sequences is constructed as follows:

```
void swapfunc(char *a, char *b, int n, int swaptype)
{   if (swaptype <=1 ) swapvector(long,a,b,n)
    else swapvector(char,a,b,n)
}
```

The following macro means swapping two long-size objects in line, and invoking the swapfunc function for other cases:

```
#define swap(a,b)                      \
    if (swaptype == 0) {               \
        long t = * (long *) (a);       \
        * (long *) (a) = * (long *) (b);   \
        * (long *) (b) = t;            \
    }                                  \
    else swapfunc(a,b,es,swaptype)
```

The following is the definitions for constants $p$, $\beta_1$ and $\beta_2$:
```
#define p 16
#define beta1 256
#define beta2 512
```

The SymPartitionSort function is an optimized Symmetry Partition Sort, which will be called by Adaptive This function sorts a partially sorted array a. Several interface parameters are defined as follows. s: the number of the initial sorted items, n : the number of items to be sorted, es : the size in bytes of each item, cmp : comparator.

```
void SymPartitionSort(char *a, int s, int n, int es, int (*cmp)(const void *,const void *))
{   char *pm,*pb,*pc,*pi,*pj;
    int i,v,vL,m,left,right,swaptype,sp,eq,ineq,rc;

    SWAPINIT(a,es);
    while(1){
        if(n < 8){  //Insertion sort on small arrays
            for (s=1; s < n; s++)
                for (pb = a+s*es; cmp(pb-es,pb) > 0; ) {
                    swap(pb,pb-es); pb-=es;
                    if(pb <= a) break;
                }
            return;
        }
        m= s<0 ? -s:s;
        if(m <= 2){ //First,middle,last items are ordered and placed 1st,2nd and last
            v = beta2 > n ? n : 63;
            pc=a+(v-1)*es;
            pm=a+es;
            swap(pm,a+(v/2)*es);
            if(cmp(a, pm) > 0) swap(a,pm);
            if((cmp(pm, pc) > 0)) {
                swap(pm,pc);
                if((cmp(a, pm) > 0)) swap(a,pm);
            }
```

```
                    left=right=1; pc-=es;
            }
            else{
                v=m > n/beta1 ? n : p*m-1;
                if(s < 0) {  //Move sorted items to left end
                    if(v<n) {left=m; s=-s;}
                    else    {left=(m+1)/2; right=m/2;}
                    swapfunc(a, a+(n-m)*es, left*es, swaptype);
                    left--;
                }
                if(s>0){
                    pb=a+m*es; pc=a+v*es;
                    if(v < n){  //Extract sampling items
                        sp=(n/v)*es; pj=pb; pi=pb;
                        for(; pi < pc; pi+=es, pj+=sp) swap(pi,pj);
                    }
                    i=right=m/2; //Right move sorted items
                    do{ pb-=es; pc-=es; swap(pb,pc); i--;} while (i);
                    left=(m-1)/2;
                }
                pm=a+left*es; pc=pm+(v-m)*es;
            }
//Fat partition begins
        pb=pi=pm+es;
        do {
            while ( (rc=cmp(pb,pm)) < 0 ) pb+=es;
            if(pb >= pc) break;
            if(rc==0){
                if(pi!=pb) swap(pb,pi);
                pi+=es; pb+=es;
                continue;
            }
            while ((rc=cmp(pc,pm)) > 0 ) pc-=es;
            if(pb >= pc) break;
            swap(pb,pc);
            if(rc==0){
                if(pi!=pb) swap(pb,pi);
                pi+=es;
            }
            pb+=es; pc-=es;
        } while (pb <= pc);
//Move equal-key items
        eq=pi-pm, ineq=pb-pi;
        if( ineq < eq) pi=pm+ineq;
        pc=pb;
        while (pm < pi ) { pc-=es; swap(pc,pm); pm+=es;}
//Fat partition ends
        vL=(pb-a)/es;
        if(right < v-vL) SymPartitionSort(pb, -right, v-vL, es, cmp);
        vL=vL-eq/es;
        if(v < n){
            if(left < vL) SymPartitionSort(a, left,vL,es,cmp);
            s=v;  //Remove tail recursion
        }
        else{
            if(left >= vL) return;
            s=left; n=vL; //Remove tail recursion
        }
    }
}
```

The Adp_SymPsort function is an Adaptive Symmetry Partition Sort supporting an interface identical to the C library function qsort. This function sorts the specified sequence into ascending order. Interface parameters are defined as follows. a : pointer to the array to be sorted, n : the number of items to be sorted, es : the size in bytes of each item, cmp : comparator.

```
#define lambda 512
#define gamma 97
void Adp_SymPsort(char *a, int n, int es, int (*cmp)(const void *,const void *))
{   char *pb,*pc,*pi,*pj;
    int swaptype,i,j,ne,rc,D_inv,left,m,Rev=0;

    SWAPINIT(a,es);
//Find 1st run
    ne = n * es;
    for (i=es; i < ne; i+=es){
        if((rc=cmp(a+i-es,a+i)) != 0 ){
            if(Rev==0) Rev= rc < 0 ? 1 : -1; //Rev=1: increasing, -1: decreasing
            else if(rc*Rev > 0) break;
        }
    }
    D_inv= Rev*(i/es);   //D_inv: difference of inversions & orders
    for(j=i+es; j < ne; j+=( gamma *es)){
        if((rc=cmp(a+j-es,a+j)) < 0) D_inv++;
        if(rc>0) D_inv--;
    }
    pb=a+i-es;
    if(abs(D_inv) > n/ lambda ) {
        if(Rev*D_inv < 0) {pb=a; Rev=-Rev;}   //If 1st run is reverse, re-find it
        pc=a+n*es; pj=pb;
        while(1){
            pj=pj+10*es; pi=pj-es;
            if(pj >= pc) break;
            while (pj < pc && Rev*cmp(pj-es, pj) <=0) pj+=es; //Find next run forward
            while (pi > pb && Rev*cmp(pi-es, pi) <=0) pi-=es; //Find next run backward
            if(pj-pi < 4*es) continue;
            if(pb!=a) {  //Find knots in 1st and 2nd run
               j=((pj-pi)/es)/2;
               m=((pb-a)/es)/4;
               if (j > m ) j=m;
               for(i=0; i<j; i++) if(Rev*cmp(pb-i*es,pi+i*es) <= 0) break;
               if(i>=j) continue;
               pb=pb+(1-i)*es; pi=pi+i*es;
            }
            // Merge two runs by moving 2nd knot to 1st knot
            if(pi!=pb) while(pi < pj ) { swap(pb,pi); pb+=es; pi+=es;}
            else pb=pj;
            pb-=es;
        }
    }
    left=(pb-a)/es+1;
    if(Rev==-1){  //If the longest run is reverse, reverse it
        pc=a;
        while(pc < pb ) {swap(pc,pb); pc+=es; pb-=es; }
    }
    if(left < n) SymPartitionSort(a, left, n, es, cmp);
}
```